\font\slbf=cmbxti10
\font\smc=cmcsc10 scaled 913
\magnification= \magstep1
\baselineskip=15pt
\hsize=6.truein
\vsize=9.truein
\def\cg{\lbrack}
\def\cd{\rbrack}
\def\cg{\char'133 }
\def\cd{\char'135 }
\def\tita#1{\goodbreak\vskip .6truecm \noindent
             {\bf #1}\nobreak\vskip .4truecm}
\def\titc#1{\goodbreak\medskip\noindent
             {\slbf #1}\nobreak\smallskip}
\def\ref(#1){{\lbrack}#1{\rbrack}}
\mathchardef\lt="313C
\mathchardef\gt="313E
\def\ovl{\overline}
\newcount\numpub
\def\aut#1*{\advance \numpub by 1
            \smallskip
            \hangindent 1.5 truecm
            \noindent \hskip 1 truecm
            \llap{\cg\the\numpub\cd\hskip.35em}
            {\smc#1}}
\def\suitaut#1*{\par
                \hangindent 1.5 truecm
                \noindent \hskip 1 truecm
                {\smc#1}}
\def\livre#1*{\enskip ---\enskip #1\ignorespaces}
\def\editeur#1*{\enskip ---\enskip {\it #1}}
\def\gras#1*{\ {\bf #1} }
\def\vep{\varepsilon}
\def\noi{\noindent}

\def\ca{{\cal C}}
\hfuzz=5pt
{\nopagenumbers
 
\null\vskip 1truecm
\centerline{\bf PHASE SPACE CHARACTERISTICS OF FRAGMENTING}
\medskip
\centerline{\bf NUCLEI DESCRIBED AS EXCITED DISORDERED SYSTEMS}
\vskip 2truecm
 
\centerline{J. Richert, D. Boos\'e, A. Lejeune\footnote{*}{On leave of absence
from the Laboratoire de Physique Nucl\'eaire Th\'eorique, Institut de Physique B5,
Universit\'e de Li\`ege, B-4000 Sart Tilman, Li\`ege, Belgium.}}
\medskip
\centerline{Laboratoire de Physique Th\'eorique, Universit\'e Louis Pasteur,}
\centerline{3, Rue de l'Universit\'e, 67084 Strasbourg Cedex, France}
\medskip
\centerline{and}
\medskip
\centerline{P. Wagner}
\medskip
\centerline{Centre de Recherches Nucl\'eaires, IN2P3/CNRS and Universit\'e
Louis Pasteur,}
\centerline{BP28, 67037 Strasbourg Cedex 2, France}
 
\vskip 3truecm
\centerline{\bf Abstract}
\bigskip
We investigate the thermodynamical content of a cellular model which describes
nuclear fragmentation as a process taking place in an excited disordered
system. The model which reproduces very well the size distribution of fragments
does not show the existence of a first order phase transition.
\vfill\eject}
 
\pageno=1
 
\tita{1. Introduction}
 
The description and understanding of nuclear fragmentation has remarkably
progressed during these last years. There remains nevertheless a long standing
question for which no completely satisfactory answer has yet been
found. It concerns the existence of an energy triggered signal which would
reveal the existence of a phase transition in nuclear matter.
\medskip
Among the advances which have come out through the confrontation of
experi\-mental measurements \ref(1-5) with theoretical investigations
one may mention the behaviour of final fragment charge distributions.
They show that the fragmentation process leads to universal properties which
can be interpreted in the framework of very simple descriptions \ref(6). In
particular, the interpretation of percolation analy\-ses points towards the
existence of a second order phase transition which takes place in an excited
disordered system \ref(7-8). More recently, experimental data \ref(9) were
used in order to relate the excitation energy to a temperature which
characterises the fragmenting system produced in peripheral collisions. The
caloric curve which comes out of these data shows a typical plateau in
temperature, with some kind of up-swing on the lower side of the energy
interval over which the effect is observed. Whether this result and its
interpretation will be definitely confirmed or not is, at present, an
open question. First, it has to be cross-checked that the temperature which
was obtained by means of a model \ref(10) is consistent with other
experimental methods aimed to fix this quantity. Up to now, this is not clear
\ref(11). Second, the concept of temperature itself makes sense only if the
system is in thermodynamical equilibrium which is tacitly supposed to be
realised, although  it can never be rigorously so. Indeed, one expects the
presence of more or less strong non-equilibrium features like collective
flow, even in peripheral kinematics. This point has to be kept in mind when
a comparison between
theoretical model calculations and experimental results is performed.
\medskip
Many different theoretical models have been proposed in order to describe
nuclear fragmentation. The specific question concerning the thermodynamic
pro\-perties of fragmenting nuclei and the possible existence of different
phases has been investigated by means of time-dependent approaches \ref(12-14)
as well as thermodynamic equilibrium models which describe the freeze-out
stage of the process in the framework of the canonical \ref(16) and the
microcanonical ensembles \ref(15). Both of these last
 calculations lead to a caloric curve
which shows the characteristic behaviour of a first order phase transition
taking place in a finite system. This may be interpreted as a support
of the experimental result, although, as already stated,
 it is not clearly established that the physical
conditions which are experimentally realised are those introduced by the models.
\medskip
In a former work \ref(7,8) we proposed a classical microscopic description in
which a fragmenting nucleus is an excited and disordered system. The approach
is able to reproduce the results of bond percolation as well as the ALADIN
data \ref(17,18) concerning charge distributions of fragments and related
observables.
This success legitimates the interest in further phase space
properties of the model. In particular, it is worthwhile to see whether it
can or cannot reproduce the thermodynamic properties predicted by the recent
experiments.
\medskip
The organisation of the present paper is the following. In section 2 we first
recall the essential features of our approach and describe the way we construct
a partition function of the system in the framework of the microcanonical
ensemble (fixed energy). We discuss the way through which we establish the
correspondence between the energy and a temperature and present the results
of our numerical investigations. In section 3 we develop an analytical version
of our approach which we confront with the numerical findings. We discuss
our results and draw conclusions in section~4.
 
\tita{2. Description of a disordered and excited system of particles and
its thermodynamical properties}
\titc{2.1 Cellular model approach to nuclear fragmentation}
 
We start with the description of the model which has already been introduced
in Refs. \ref(7,8). We consider a finite compact volume $V$ in 3-dimensional
 space,
in practice a cube. This volume is divided into $A$ equal cubic cells of linear
dimension $d$. Each cell is occupied by a classical particle, i.e. $V = Ad^3$.
Each particle $i$ is characterized by its phase space coordinates, i.e. its
position $\vec r_i$ with respect to the centre of mass of the system and its
linear momentum $\vec p_i$. The coordinates ($x_i, y_i, z_i$) are chosen as
$x_i = x_{i_0} + \vep d/2$ and similarly for $y_i$ and $z_i$. The coordinate
$\vec r_{i_0} (x_{i_0}, y_{i_0}, z_{i_0}$) corresponds to the centre of the
cell and $\vep$ is a random number drawn from a distribution which, in
practice, is chosen to be uniform in the interval [0,1]. The momenta $\vec p_i$
are drawn randomly and uniformly in a sphere of arbitrary large radius
$p_{\max}$. We introduce
the maximum kinetic energy of a particle as $E_{\max} = p^2_{\max}/2m$.
\medskip
The particles interact by means of two-body interactions. In practice we
introduce a strong short range nuclear potential. For reasons of consistency
with former calculations \ref(8) we essentially use the interaction of
Ref. \ref(19) which does not explicitly differentiate protons and neutrons.
If we select another potential in the sequel we shall mention it in due place.
We choose part of the nucleons to be protons which we distribute uniformly
in cells over the whole space and implement the
Coulomb potential which acts between charged particles. The potential energy
of the system is obtained by adding up the interaction energy between the
particles. We have also considered the case where the nuclear interaction is
restricted to nearest neighbours. This approximation changes
absolute values of energies but does not affect the conclusions which will be
drawn below.
\medskip
Former use of this classical approach has nicely proven that the present
des\-cription of excited fragmenting nuclear systems is able to reproduce the
fragment size distributions and related observables which are obtained in the
framework of ordinary bond percolation \ref(7,8). The calculations show that
the results are very robust with respect to the explicit choice of the
interaction, the shape of the external surface of the system and even the
criteria which are used in order to define bound systems of particles, i.e.
fragments.
 
\titc{2.2 Thermodynamical aspects}
 
\noi
{\it 2.2.1 Motivations}
\medskip
The agreement between percolation calculations which are generic for the
des\-cription of disordered systems and experimental results suggests a possible
interpretation of fragmentation as a critical phenomenon, more precisely a
second order phase transition in presence of finite size
effects since it takes place in a small system. A recent interpretation of
experimental data raises a new question. Indeed, the determination of the
temperature of the excited fragmenting system and the know\-ledge of the
excitation energy of projectile-like nuclei formed in peripheral collisions
lead to a caloric curve which exhibits the characteristic plateau corresponding
to latent heat generation. This effect which is seen in infinite systems
undergoing a first order phase transition is interpreted as a liquid-gas-like
transition \ref(9). This result which as already said,
 is still under critical investigation
and its interpretation led us naturally to the question as to whether
our approach can reproduce such an effect or not. This is the point we want
to present and discuss in the sequel.
\bigskip
\noi
{\it 2.2.2 Generation of microcanonical configurations}
\medskip
The total number of configurations which can be generated for a fixed energy
$E$, number of particles $A$ and volume $V = Ad^3$ is designated by
$Z\,(E, A, d)$, the microcanonical partition function of the system when it
is in thermodynamical equilibrium.
\medskip
In practice we construct this quantity numerically in the following way. We
generate $N$ configurations with our cellular model as described above
and calculate the potential energy of the system by summing up the two-body
interaction energy which acts between all pairs of particles of the system,
i.e. the nuclear and the Coulomb contributions. The number $Z$ of charged
particles is such that $Z/A$ corresponds to nuclei energetically located
in the valley of stability. Doing this way we obtain $N$ configurations with
potential energy $\lbrace E^{(k)}_{pot}, k = 1,\ldots, N\rbrace$.
\medskip
In a second step we generate a set of configurations for which we determine
the kinetic energy content. For each configuration we choose randomly
$p_{\max}$ (see above) out of a uniform distribution in an interval
[$0, P_{up}$] where $P_{up}$ is an upper limit for the momenta which we fix
arbitrarily. We then draw randomly, for fixed $p_{\max}$, an ensemble of
momenta $\lbrace \vert \vec p_i\vert, i = 1,\ldots, A\rbrace$ in the Fermi
sphere of radius $p_{\max}$ as already described above. Proceeding this way
we generate a set of events with total kinetic energy
$\lbrace E^{(k)}_{kin}, k = 1,\ldots, N\rbrace$. The number of events is
generally of the order of $10^6$.
\medskip
We fix then the total energy of the system to lie in bins
[$E, E + \Delta E$] and
count the number of configurations which correspond to a total energy
$E = E_{kin} + E_{pot}$ lying in this bin by means of the combination of the
kinetic and potential energy configurations which were generated as described
above. The number of configurations which are found gives the partition
function $Z\,(E, A, d$). This is done for all energies which the system can
reach, from a minimum value $E_{\min}$ to the maximum which is
$E_{up} = A\cdot P_{up}^2/2m$.  $E_{up}$ is necessarily
a finite quantity because it is necessary to define a finite upper bound in
numerical simulations. This leads to the fact that $Z\,(E, A, d)$ may start
to
decrease for the higher values of $E$. This is an unphysical effect. Hence we
must restrict the further use of the partition function to an interval of
energy which does not take its high energy tail into account.
\bigskip
\noi
{\it 2.2.3 Thermodynamical analysis of the disordered system. Results}
\medskip
The knowledge of the microcanonical partition function leads to the entropy
$S\,(E, A, d)$ and the temperature through
$$T^{-1} = (\partial S/\partial E)_V\eqno(1)$$
 
\noi
which establishes the relationship between $T$ and $E$, i.e. the so called
caloric curve.
\medskip
We have determined $Z\,(E, A, d)$ for different values of $d$, i.e. different
total volumes. In order to work out the derivative of $S = \ln Z$ we have
introduced an ana\-lytical parametrisation of $S$ over the whole range of
values in terms of Chebyshev polynomials of different orders. Using
statistically
significant samples of configurations (of the order of $10^6$) it is possible
to approximate $S$ with an excellent precision ($\chi^2 \leq 1$).
Fig. 1a shows the behaviour of $Z$ obtained by means of the numerical
simulations described above and the result of the analytical fit which is
practically indistinguishable from the numerical result. The corresponding
caloric curve is represented in  Fig. 1b. There is no sign for the existence
of a plateau in the temperature which would signal a first order phase
transition.
\medskip
In Fig. 2 we present the behaviour of $(\partial S/\partial E)^{-1}_V$ for
different values of $V$ which correspond to values of $d$ in the interval
[1.8,~4]$\,$fm (average densities $\rho$ in the interval
[0.016,~0.17]$\,\rm fm^{-3}$).
One observes a strong dispersion of the curves
as a function of the energy which indicates that this quantity is sensitive
to the volume in which the system moves. The large negative energy thresholds
come from the fact that the nucleon potential we chose \ref(19) and which
was also used in Refs. \ref(7,8) overbinds heavy systems (here $A=216)$.
In our former study of fragment size
distributions \ref(7,8) we mixed up events which correspond to systems of
different sizes. Since, for different experimental events, the nucleus may
break up into fragments at different densities it is tempting to collect in a
uniform way events corresponding to different values of $d$. This is what
we did in order to construct an ``entropy" which is averaged over different
values of $d$, $\lt S\gt_d$. In Fig. 2 we show the behaviour of
($\partial \lt S\gt_d /\partial E)^{-1}_V$ (dotted line).
Again, this quantity shows a monotonous rise with the energy.
\medskip
We also used the potential defined in Ref. \ref(20) which differentiates protons
and neutrons and hence leads to a quite different energy dependence of the system.
As it can be seen in Fig. 3 one finds the same behaviour of the observables
as in the former case.
\bigskip
\noi
{\it 2.2.4 Comments}
\medskip
We comment and discuss now the results obtained above. First, the fact that
we take together events corresponding to different volumes violates (1) since
$\partial S/\partial E$ must be calculated for fixed volume. The only way out
is to say that the quantity which corresponds to
($\partial\lt S\gt_d/\partial E)^{-1}$ is some effective temperature. It is
not clear whether it could be put in connection with temperatures related to the
experimental situation where the fragmentation may go on in a variable volume.
\medskip
Second, it is interesting to confront the behaviour of the partition sum
$Z\,(E, A, d)$ with the multiplicity as a function of increasing energy. This
is shown in Fig. 4. As expected there exists a clear correspondence between
the increase of the number of configurations and the increase of the number of
fragments and particles over the whole energy range. However this connection
is not related to a phase transition, as already stated above (no exponential
increase of $Z\,(E, A, d)$ over a finite energy interval in which one would
observe a coexistence of fragments and particles).
\medskip
 
Last, thermodynamical equilibrium which is a prerequisite
for the use of (1). There is no reason why all the configurations we generate
should correspond to such a situation since we do not implement a Metropolis
Monte Carlo type
algorithm in order to select those configurations which correspond
to equilibrium.
We try now to eliminate the problem concerning thermodynamical equilibrium
by implementing an analytical approach.
 
\tita{3. Analytic approach to the description of an excited disordered
system}
As previously discussed,
the numerical simulations which were performed above do
not necessarily generate an equilibrated system. It raises the questions
whether the outcome of our model would be different if this property would
be satisfied. This should be done by implementing some Metropolis Monte Carlo
algorithm. Such calculations may be done in the future. Here we try to
work out an analytical model which guarantees thermodynamic equilibrium.
Furthermore, its geometric and dynamical properties are very close to those
of the system which has been investigated above. We want to check the effects
of possible deviations from equilibrium which is not realised in the
 calculations of section 2.
 
\titc{3.1 Derivation of the canonical partition function}
 
We construct the canonical partition function of a set of $A$ classical particles
in thermodynamic equilibrium at a temperature	
$T = \beta^{-1}$. The particles are distributed in $A$ identical cells of
linear dimension $d$. As before, there is one particle in each cell. Then
 
$$\widetilde Z\,(\beta, A, d) = \sum_{\ca\,(\lbrace \vec r_i, \vec p_i\rbrace)}
e^{-\beta H(A, d)}\eqno(2)$$
$$=\prod^A_{i=1}~\int_{V_{p_i}} d\vec p_i \int_{V_{r_i}} d\vec r_i\, e^{-\beta
H (\lbrace \vec r_i, \vec p_i\rbrace)}$$
with
$$H\,(\lbrace \vec r_i, \vec p_i\rbrace) = \sum^A_{i=1}~{p_i^2\over 2m} +
\sum_{i\lt j}\,V(\vert \vec r_i - \vec r_j\vert)$$
\bigskip
\noi
$V_{p_i}$ and $V_{r_i}$ are the volumes occupied by a particle in a phase
space cell.
We consider first a 1-dimensional system for which cells are segments of length
$d$. The potential $V$ is defined in terms of a truncated harmonic oscillator
which acts between neighbouring particles $i$ and $j$ located at $x_i, x_j$
such that
$$\eqalign{
V&= \infty\hskip 3.5truecm ~\vert x_i-x_j\vert \lt r_c\cr
&=V_0 + C\,(x_i - x_j)^2\quad ~~~r_c \leq \vert x_i - x_j\vert \leq r_0\cr
&=0\hskip 3.5truecm ~~\,\vert x_i - x_j \vert \gt r_0\cr}$$
\medskip
The parameter $r_c$ introduces a hard core. The effective range of the
interaction $r_0$ is chosen to be of the order of $2d$. The strength $V_0$ is
negative and $C = -V_0/r_0^2$.
\medskip
The potential $V$ is aimed to simulate a realistic short range nuclear
potential which is attractive and strongly repulsive at short distances.
\medskip
The model can be easily extended to 3 dimensions (see below). It is expected
to be qualitatively close to the more realistic potentials used in numerical
simulations. The long range Coulomb interaction is not taken into account.
From our numerical experience this should not alter our conclusions.
 
\titc{3.2 Kinetic and potential contributions to the partition function}
 
The kinetic contribution is trivially given by
$$\widetilde Z_{kin} (\beta, A) = \prod^A_{i=1}~ \int^{+\infty}_{-\infty}\,
d\,p_i\exp\,(-\beta p_i^2/2m) = (2\pi m/\beta)^{A/2}\eqno(3)$$
\medskip
The potential contribution can also be worked out explicitly.
The potential energy reads
$$V = AV_0 + C\left\lbrack (x_1-x_2)^2 + \ldots + (x_{A-1} - x_A)^2 + (x_A -
x_1)^2\right\rbrack$$
 
\noi
We impose periodic boundary conditions
$$x_{A+1} = x_1$$
 
\noi
and introduce new coordinates
$$x_i = x_i^0 + \delta x_i\quad;\quad\delta x_{A+1} = \delta x_1$$
 
\noi
where $x_i^0$ defines the centre of the interval $i$.
\medskip
\noi
Then
$$V = AV_0 + ACd^2 + C\,\sum^A_{i=1}\,(\delta x_{i+1} - \delta x_i)^2 +
2Cd\,\sum^A_{i=1}\,(\delta x_{i+1} - \delta x_i)$$
\medskip
The periodic boundary condition eliminates the last term on the right hand
side. Hence the potential contribution to the partition function reads
$$\widetilde Z_{pot}(\beta, A, d) = \exp\left\lbrack-\beta A(V_0+Cd^2)\right
\rbrack\cdot \prod^A_{k=1}\int^b_{-b} d\delta x_k\exp\left\lbrack-\beta C
\sum^A_{i=1}(\delta x_{i+1} - \delta x_i)^2\right\rbrack$$
where $\lbrack -b, +b\rbrack$ with $b=d/2-r_c$ is the range over which
$\delta x_i$ can vary in the interval~$i$.
\medskip
The exponent of the integrant in $\widetilde Z_{pot}$ is a quadratic form
which corresponds to a matrix with $A$ equal diagonal elements $2\beta C$,
two symmetrical subdiagonals with equal elements $ - \beta C$ and two
elements $-\beta C$,
one in the upper right corner and one in the lower left corner. This
``tight-binding"
matrix can be explicitly diagonalised and the eigenvalues are
given by
$$\widetilde\lambda_i = 2\beta C\left(1-\cos~\left({2\pi i\over A}\right)\right)
=\beta C\lambda_i\quad (i=1,\ldots, A)$$
\medskip
The lowest eigenvalue corresponds to $\widetilde\lambda_A = 0$. The other
eigenvalues are degenerate of order 2, $\widetilde\lambda_{A-i} =
\widetilde\lambda_i$. If $A$ is even, the
eigenvalue $\widetilde\lambda_{A/2} = 4\beta C$ is the largest and it is non
degenerate.
\medskip
It is also possible to work out the expression of the eigenvectors and to
obtain the new boundaries in the new variables $\{\delta X_i\}$ corresponding
to the rotated system and associated with the eigenvalues
$\{\widetilde\lambda_i\}$.
\medskip
For $A$ odd one gets
$$-b\,\sqrt A~~\leq~~\delta X_1~~\leq +~b\,\sqrt A\qquad{\rm corresponding~to}~~
\lambda_A$$
and
$$\eqalign{
&-b_i~~~\leq~~~ \delta X_{2i}~~~~~\leq~+~b_i\cr
&-c_i~~~\leq~~~ \delta X_{2i+1}~~\leq~+~c_i\cr}$$
where
$$\eqalign{
b_i &= b~\sqrt{{2\over A}}\, \left(1+2\sum^{{A-1\over 2}}_{k=1} \left\vert
\cos \left({2\pi k\over A}\cdot i\right)\right\vert\right)\cr
\cr
c_i &= b~\sqrt{{2\over A}}\, \left(1+2 \sum^{{A-1\over 2}}_{k=1} \left\vert
\sin \left({2\pi k\over A}\cdot i\right)\right\vert\right)\cr}$$
which correspond to the degenerate eigenvalue $\lambda_i$.
\medskip
For $A$ even the boundaries are
$$-b~\sqrt A~\leq~ \delta X_{1,2}~ \leq~+~b~\sqrt A$$
 
\noi
corresponding to $\lambda_A$ and $\lambda_{A/2}$
\smallskip
\noi
and
$$\eqalign{
&-b_i~~~\leq~~~ \delta X_{2i+1}~~\leq~+~b_i\cr
&-c_i~~~\leq~~~ \delta X_{2i+2}~~\leq~+~c_i\cr}$$
corresponding to the degenerate eigenvalues $\lambda_i$. Here $b_i$ and $c_i$
are the same as those given above, except that the upper limit
of the sums corresponds now to $k = A/2 - 1$.
\medskip
As a consequence of these developments the expression of $\widetilde Z_{pot}$
can be explicitly integrated for finite $A$. For $A$ odd one gets
$$\eqalignno{
&\widetilde Z_{pot}^{(o)} (\beta, A, d) = \exp\left\lbrack-\beta A(V_0 + Cd^2)
\right\rbrack\cdot 2b\,A^{1/2}\cdot (\pi/\beta C)^{(A-1)/2}\cr
\cr
&\cdot \left(\prod^{(A-1)/2}_{i=1} \lambda_i\right)^{-1}\cdot
\prod^{(A-1)/2}_{i=1} \left\lbrace{\rm erf} \left\lbrack
(\beta C\lambda_i)^{1/2} b_i\right\rbrack\cdot{\rm erf}\left\lbrack
(\beta C\lambda_i)^{1/2} c_i\right\rbrack\right\rbrace&(\rm 4a)\cr}$$
\medskip
\noi
and for $A$ even the expression reads
$$\eqalignno{
&\widetilde Z_{pot}^{(e)} (\beta, A, d) = \exp\left\lbrack-\beta A(V_0 + Cd^2)
\right\rbrack\cdot 2b\,A^{1/2}\cdot (\pi/\beta C\lambda_{A/2})^{1/2}\cdot
(\pi/\beta C)^{(A-2)}\cr
\cr
&\cdot \left(\prod^{(A-2)/2}_{i=1} \lambda_i\right)^{-1}\cdot
\prod^{A/2-1}_{i=1} \left\lbrace{\rm erf} \left\lbrack
(\beta C\lambda_i)^{1/2} b_i\right\rbrack\cdot{\rm erf}\left\lbrack
(\beta C\lambda_i)^{1/2} c_i\right\rbrack\right\rbrace&(\rm 4b)\cr}$$
\medskip
\noi
where erf$\,(x$) is the error function and $\lambda_i=
\widetilde\lambda_i/\beta C$
($i=1,\,\ldots,\, A$).
 
\titc{3.3 Approximate expressions}
 
It is now possible to write down $\widetilde Z\,(\beta, A, d)$ as the
product of $\widetilde Z_{kin}(\beta, A)$ given by (3) and
$\widetilde Z_{pot}(\beta, A, d)$ given by (4a) or (4b). This is however
of no practical use since it would not allow us to get an explicit expression
of the microcanonical partition function $Z\,(E, A, d)$ which we would like
to confront with our simulation results.
\medskip
For this reason we introduce an approximation in our calculation of
$\widetilde Z_{pot}$. In practice, we replace the rotated hypercube which
defines the volume in which the $A-1$ rotated coordinates $\lbrace \delta X_i
\rbrace$ vary by a hypersphere whose radius $R$ is defined in such a way that
its volume is the same as that of the original hypercube of dimension $A-1$
(the coordinate corresponding to $\lambda_A = 0$ is not taken into account).
\medskip
This procedure leads to
$$R=\lbrack V\cdot \Gamma\,(A + 1)/2\rbrack ^{1/(A-1)}/\pi^{1/2}\eqno(5)$$
where $V = (2b)^{(A-1)}\cdot\left(\prod\limits_{k=1}^{(A-1)/2}\lambda_k
\right)$ for $A$ odd.
\medskip \noi
If we introduce this approximation we get the partition function
$$\eqalignno{
&\ovl Z_{pot}^{\,(o)} (\beta, A, d) = \exp\left\lbrack-\beta A(V_0 + Cd^2)
\right\rbrack\, 2b\,A^{1/2}\cdot (\pi/\beta C)^{(A-1)/2}\cr
\cr
&\cdot \left(\prod^{(A-1)/2}_{i=1} \lambda_i\right)^{-1}\cdot
\left\lbrack 1-\exp(-\beta CR^2)\,\sum^{(A-3)/2}_{m=0}\,
{(\beta CR^2)^m\over m!}\right\rbrack&(\rm 6a)\cr}$$
\medskip \noi
for the case where $A$ is odd, and
$$\eqalignno{
&\ovl Z_{pot}^{\,(e)} (\beta, A, d) = \exp\left\lbrack-\beta A(V_0 + Cd^2)
\right\rbrack\cdot 2b\,A^{1/2}\cdot (\pi/\beta C\lambda_{A/2})^{1/2}
\cdot (\pi/\beta C)^{(A-2)}\cr
\cr
&\cdot{\rm erf} \left\lbrack b\,A^{1/2} (\beta C\lambda_{A/2})^{1/2}\right\rbrack
\cdot \left(\prod^{(A-2)/2}_{i=1} \lambda_i\right)^{-1}\cdot
\left\lbrack 1-\exp(-\beta CR^2)\,\sum^{(A-4)/2}_{m=0}\,
{(\beta CR^2)^m\over m!}\right\rbrack\cr
&&(\rm 6b)\cr}$$
for the case where $A$ is even. In (6b) $R$ is defined by (5), but $A-1$ is
replaced by $A-2$.
\medskip
The exact and approximate expressions $\widetilde Z_{pot}$ and
$\ovl Z_{pot}$ are close and it is easy to guess that their behaviour
as a function of $\beta$ is qualitatively similar. We have checked this point
by means of explicit numerical calculations in which we compare the product
of error functions appearing in (4a) and (4b) with the $R-$ dependent terms
in (6a) and (6b). Calculations done for different values of $A$ and different
choices of parameters are shown in Fig. 5 for the quantities by which
$\widetilde Z^{(o)}$ and $\ovl Z{\,}^{(o)}$ given by (4a) and (6a) differ (last
terms of the expressions). The results show that these quantities have the
same monotonous behaviour, with the same
limit as $\beta$ goes to infinity. In both cases the derivatives of the curves
go to zero as $\beta$ goes to zero.

\titc{3.4 Microcanonical partition function}
 
We use $\ovl Z$ in order to construct the microcanonical partition function
by means of an inverse Laplace transform
$$Z\,(E, A, d) = {1\over 2\pi i} \int^{c+i\,\infty}_{c-i\,\infty}\,d\beta\,
e^{\,\beta E}~\ovl Z (\beta, A, d)\eqno(7)$$
\noi
Because of the explicit dependence of $\ovl Z$ on $\beta$ it is possible to
work out $Z$ explicitly.
\medskip
It should be noticed here that inverse Laplace transforms can be highly
unstable quantities \ref(21,22). This fact is particularly troublesome
when $Z{(\beta)}$ is numerically determined since any deviation of
$Z{(\beta)}$ can produce important deviations of $Z(E)$ from its real
value. In our case $\widetilde Z{(\beta)}$ is known analytically. We
believe that, even though $\ovl Z(E)$ we obtain from this analytic
expression is not the same as the exact one, $\ovl Z(E)$ reflects the
qualitative behaviour of the exact partition function $\widetilde Z(E)$
because of the analytical behaviour of both functions
$\widetilde Z{(\beta)}$ and $\ovl Z{(\beta)}$ (see above and Fig.~5).
This is sufficient in order to conclude confidently about the qualitative
behaviour of $E$ vs. $T$ (see below and Figs.~6 and 7).
\medskip
Using (6a) it is straightforward to show that
$$\eqalignno{
&Z\,(E, A, d) = (2\pi m)^{A/2} (\pi/C)^{(A-1)/2}
\cdot 2bA^{1/2}\cdot \left(\prod^{A-1}_{k=1}
\lambda_k^{1/2}\right)^{-1}\cr
\cr
&\cdot\left\lbrace {\lbrack (E-A (V_0 + Cd^2)\rbrack^{A-3/2}\over
\Gamma\,(A-1/2)}
\cdot \Theta\left\lbrack E - A (V_0 + Cd^2)\right\rbrack\right.\cr
\cr
&-\sum^{(A-3)/2}_{m=0}\,{C^m R^{2m}\over m!\Gamma(A-m-1/2)}\,
\left\lbrack E-A(V_0 + Cd^2)-CR^2\right\rbrack^{A-m-3/2}\cr
\cr
&\qquad\qquad\cdot\Theta\left\lbrack E-A(V_0+Cd^2)-CR^2\right\rbrack
&(8)\cr}$$
\medskip
\noi
where $R$ has been defined above. This expression shows explicitly that
$Z\,(E, A, d)$ has an algebraic dependence on $E$ starting from a threshold
energy $E = A\,(V_0 + Cd^2)$. There is no sign for an exponential behaviour
which would indicate the existence of a plateau in the ($T = \beta^{-1}, E)$
representation. The fact that $Z$ is a monotonic increasing algebraic
function of $E$
is confirmed by numerical tests of expression (8), see Fig. 6. This confirms
and comforts the results obtained through simulations in section 2.
\medskip
The analytic results can easily be extended to a 3-dimensional system since,
because of the choice of the two-body potential, the potential contribution
factorises in the 3 space dimensions ($x, y, z)$. Hence the expressions are
formally similar to those obtained in the 1-dimensional case, except for the
eigenvalue spectrum $\lbrace \lambda_k; k = 1,\ldots, A\rbrace$ which is
affected by the dimensionality of the total space because, in the
nearest-neighbour interaction approximation the
coordinance goes over from 2 in the 1-dimensional case to 6 in the
3-dimensional one. The partition function reads now explicitly
$$\eqalignno{
&Z\,(E, A, d)=(2\pi m)^{3A/2}\cdot(2bA^{1/2})^3\cdot(\pi/C)^{(A-1)/2}
\left(\prod^{A-1}_{i,j,k=1}\lambda_{i,j,k}^{1/2}\right)^{-3}\cr
\cr
&\cdot\left\lbrace \left\lbrack E-A (2V_0 + 6Cd^2)\right\rbrack^{3A-5/2}
\cdot \Theta\left\lbrack E - A (2V_0 + 6Cd^2)\right\rbrack/\Gamma
(3A-3/2)\right.\cr
\cr
&-3\sum^{(A-3)/2}_{m=0}\,{C^m R^{2m}\over m!\Gamma(3A-3/2-m)}\,\left\lbrack
E-A(2V_0 + 6Cd^2)-CR^2\right\rbrack^{3A-m-5/2}\cr
\cr
&\qquad\qquad\cdot\Theta\left\lbrack
E-A(2V_0+6Cd^2)-CR^2\right\rbrack\cr
\cr
&+3\sum^{(A-3)/2}_{m,m'=0}\,{C^{m+m'}R^{2(m+m')}\over m!m'!\Gamma
(3A-3/2-m-m')}\cr
\cr
&\qquad\qquad
\cdot\left\lbrack E-A(2V_0+6Cd^2)-2CR^2\right\rbrack^{3A-m-m'-5/2}\cr
\cr
&\qquad\qquad\cdot \Theta\left\lbrack E-A(2V_0+6Cd^2)-2CR^2\right
\rbrack\cr
\cr
&-\sum^{(A-3)/2}_{m,m',m''=0}\,{C^{m+m'+m''}R^{2(m+m'+m'')}\over
m!m'!m''!\Gamma(3A- 3/2-m-m'-m'')}\cr
\cr
&\qquad\qquad\cdot\left\lbrack E-A(2V_0+6Cd^2)
-3CR^2\right\rbrack^{3A-m-m'-m''-5/2}\cr
\cr
&\qquad\qquad\cdot \Theta\left\lbrack E-A(2V_0+6Cd^2)-3CR^2\right
\rbrack&(9)\cr}$$
\medskip
\noi
Here $R$ is given by the same expression as (5).
\medskip\noi
The eigenvalues are
$$\lambda_{i,j,k} = 2 \left(3-\cos{2\pi i\over N}-\cos{2\pi j\over N}-
\cos{2\pi k\over N}\right)$$
where ($i,\,j,\,k=1,\ldots, N)$ and $N=A^{1/3}$.
 
\medskip\noi
The correlation between $E$ and $T$ is shown in figure~7. It is similar to
the one obtained in the one-dimensional case. In practice, in both cases,
the relation between $E$ and $T$ looks like that of an ideal classical gas
$T=3/2E$, but with a different slope because of the presence of an
interaction.
 
\tita{4. Discussion and conclusions}
 
We have worked out the properties of an excited system of nucleons with
 respect to its energy.
We introduced a classical model which is aimed to describe a fragmenting
nucleus considered as a disordered system.
In this model the nucleons occupy finite phase space cells. We have shown
in former work that this approach reproduces very successfully the fragment
size distributions and related observables predicted by bond
percolation models and experimentally observed, hence the
properties related to a second order phase transition corrected for finite
size effects.
\medskip
Our interest in the thermodynamic properties of the system was triggered by
recent experimental results which indicate that there may exist a first
order phase transition in the fragmentation process. The coexistence of
phase transtitions of first and second order could be explained by the
fact that first order transitions become second order transitions at the
critical point. The universality properties of the fragment size
distributions may reflect the existence of events which lie at the
critical point and in its neighbourhood.
We used our model in
order to construct the microcanonical partition function, taking into account
the short range nuclear and long range
Coulomb interactions between all the nucleons.
It comes out of our calculations that we do not find the plateau of constant
temperature in the caloric curve which characterises a first order transition.
This is the case for numerical simulations with different nuclear interactions
and also for a simplified analytical approach which leads to a steady
increase of the
temperature with increasing energy. In numerical simulations we faced the
problem that part of the generated configurations do not correspond to thermal
equilibrium, since we did not select them by means of an appropriate
 Metropolis Monte Carlo procedure.
This is the reason for which we implemented an analytical approach which
implies thermodynamical equilibrium by construction. In order to ensure
an analytical expression for the partition function which shows explicitly
the behaviour of the caloric curve, we introduced a simplified two-body
potential. This potential reproduces the qualitative behaviour of the
realistic interaction used in numerical simulations.
 
\medskip
The reason for the failure of the present model with respect to the presence
of a phase transition are not clear. We suspect that the concept of cells
in space introduces a rigid structure which leads to a system closer to a
disordered crystal than to a liquid which may vaporize when the excitation
energy is high enough. One may of course relax this constraint and possibly,
at the same time, select those configurations which correspond to equilibrium.
But then the new approach should also reproduce the fragment size distributions
which are so nicely obtained in our present approach.
\smallskip
It is our aim to investigate this point in the future and confront it with
the outcome of cellular automata techniques which introduce a dynamical
description through an explicit time-dependence.
\bigskip
\titc{Aknowledgments}
The authors would like to thank Prof. D. Gross for fruitful comments and
advices. One of us (J.R.) aknowledges interesting discussions with
A. Bonasera, J. Polonyi and P. Simon and the help of P. Oswald for the
completion of numerical calculations.
\vfill\eject
 
\noi
{\bf References}
\bigskip
 
\aut J. Hubele et al.*
\editeur Phys. Rev.* \gras C46* (1992) R1577
\smallskip
 
\aut U. Lynen et al.*
\editeur Nucl. Phys.* \gras A545* (1992) 329c
\smallskip
 
\aut P. Creutz et al.*
\editeur Nucl. Phys.* \gras A556* (1993) 672
\smallskip
 
\aut P. D\'esesquelles et al.*
\editeur Phys. Rev.* \gras C48* (1993) 1828
\smallskip
 
\aut A. Sch\"uttauf et al.*
\livre GSI preprint 96-26, June 1996*
\smallskip
 
\aut X. Campi*
\editeur J. Phys. A : Math. Gen.* \gras 19* (1986) L917 ; {\it Phys. Lett.}
\gras B208* (1988) 351
\suitaut ~X. Campi, H. Krivine*
\livre Contribution to the International Workshop on Dynamical Features
of Nuclei and Finite Fermi Systems*, Sitges, Spain (1993) (World Scientific,
Singapore)
\smallskip
 
\aut B. Elattari, J. Richert, P. Wagner, Y.M. Zheng*
\editeur Phys. Lett.* \gras B356* (1995) 181
\smallskip
 
\aut B. Elattari, J. Richert, P. Wagner, Y.M. Zheng*
\editeur Nucl. Phys.* \gras A592* (1995) 385
\smallskip
 
\aut J. Pochodzalla et al.*
\editeur Phys. Rev. Lett.* \gras 75* (1995) 1040
\smallskip
 
\aut S. Albergo, S. Costa, E. Costanza, A. Rubbins*
\editeur Nuovo Cimento* \gras A89* (1985) 1
\smallskip
 
\aut X. Campi, H. Krivine, E. Plagnol*
\livre ``Remarks on a determination of the nuclear caloric curve"*,
preprint IPNO/TH 96-17
\smallskip
 
\aut P. Finocchiaro, M. Belkacem, J. Kubo, V. Latora, A. Bonasera*
\editeur Nucl. Phys.* \gras A600* (1996) 236
\smallskip
 
\aut S. Pratt*
\editeur Phys. Lett.* \gras B349* (1995) 261
\smallskip
 
\aut A. Guarnera, M. Colonna, Ph. Chomaz*
\editeur Phys. Lett.* \gras B373* (1996) 267 and refs. quoted therein
\smallskip
 
\aut D.H.E. Gross*
\editeur J. Phys. Soc. Japan* \gras 54* (1985) 392 ; {\it Rep. Prog. Phys.}
\gras 53* (1990) 605
\smallskip
 
\aut H.W. Barz, J.P. Bondorf, R. Donangelo, H. Schulz*
\editeur Phys. Lett.* \gras B169* (1986) 318
\smallskip
 
\aut Y.M. Zheng, J. Richert, P. Wagner*
\editeur Chinese J. Nucl. Phys.* \gras 17* (1995) 215
\smallskip
 
\aut Y.M. Zheng, J. Richert, P. Wagner*
\editeur J. Phys.* \gras G22* (1996) 505
\smallskip
 
\aut L. Wilets, E.M. Henley, M. Kraft, A.D. McKellar*
\editeur Nucl. Phys.* \gras A282* (1977) 341
\smallskip
 
\aut R.J. Lenk, T.J. Schlagel, V.R. Pandharipande*
\editeur Phys. Rev.* \gras C43* (1990) 372
\smallskip
 
\aut R.W. Gerling, A. H\"uller*
\editeur Z. Phys.* \gras B90* (1993) 207
\smallskip
 
\aut G. Arfken* {---} Mathematical methods for physicists,
New York, London, Academic Press 1985
\smallskip
 \vfill\eject

  \centerline{\bf Figure captions}
\vskip 3truecm
\itemitem{Fig. 1a:} $S(E)=\ell n\,Z(E,A,d)$ as a function of $E/A$ for
$A=216$ and $d=3\,$fm.
\vskip 1truecm
 
\itemitem{Fig. 1b:} $(\partial S/\partial E)_V^{-1}$  as a function of $E/A$
($V=Ad^3$ is the volume of the system).
\vskip 1truecm
 
\itemitem{Fig. 2:} $(\partial S/\partial E)_V^{-1}$ as a function of $E/A$
for different values of $d$ ($V=Ad^3$is the volume of the system)
; the dashed line corresponds to the case
where $S$ is averaged over different values of $d$ (see text).
\vskip 1truecm
 
\itemitem{Fig. 3:} $S(E)$ and $(\partial S/\partial E)_V^{-1}$ as a
function of $E/A$ for $d=3\,$fm with the interaction taken from ref.~[20].
\vskip 1truecm
 
\itemitem{Fig. 4:} Evolution of the partition $Z(E)$ and fragment
multiplicities as a function of $E/A$. See text. $Z(E)$ is represented
in arbitrary units.
\vskip 1truecm
\itemitem{Fig. 5:} Comparison between the exact (4a) and approximate (6a)
expressions of the canonical partition function. The quantities which are
represented correspond to the product of error functions in (4a) and the
term in brackets in (6a). See text.
\vskip 1truecm
\itemitem{Fig. 6:} $s(E)=\ell n\,Z(E,A,d)/A$ and temperature $T$
as a function of $E/A$ for  $d=3\,$fm and $A=125$ (1 dimensional system).
See text. The parameters are the same as in Fig.~5.
 \vskip 1truecm
\itemitem{Fig. 7:} Same as Fig.~6 for a 3 dimensional system $(A=5^3)$.
 
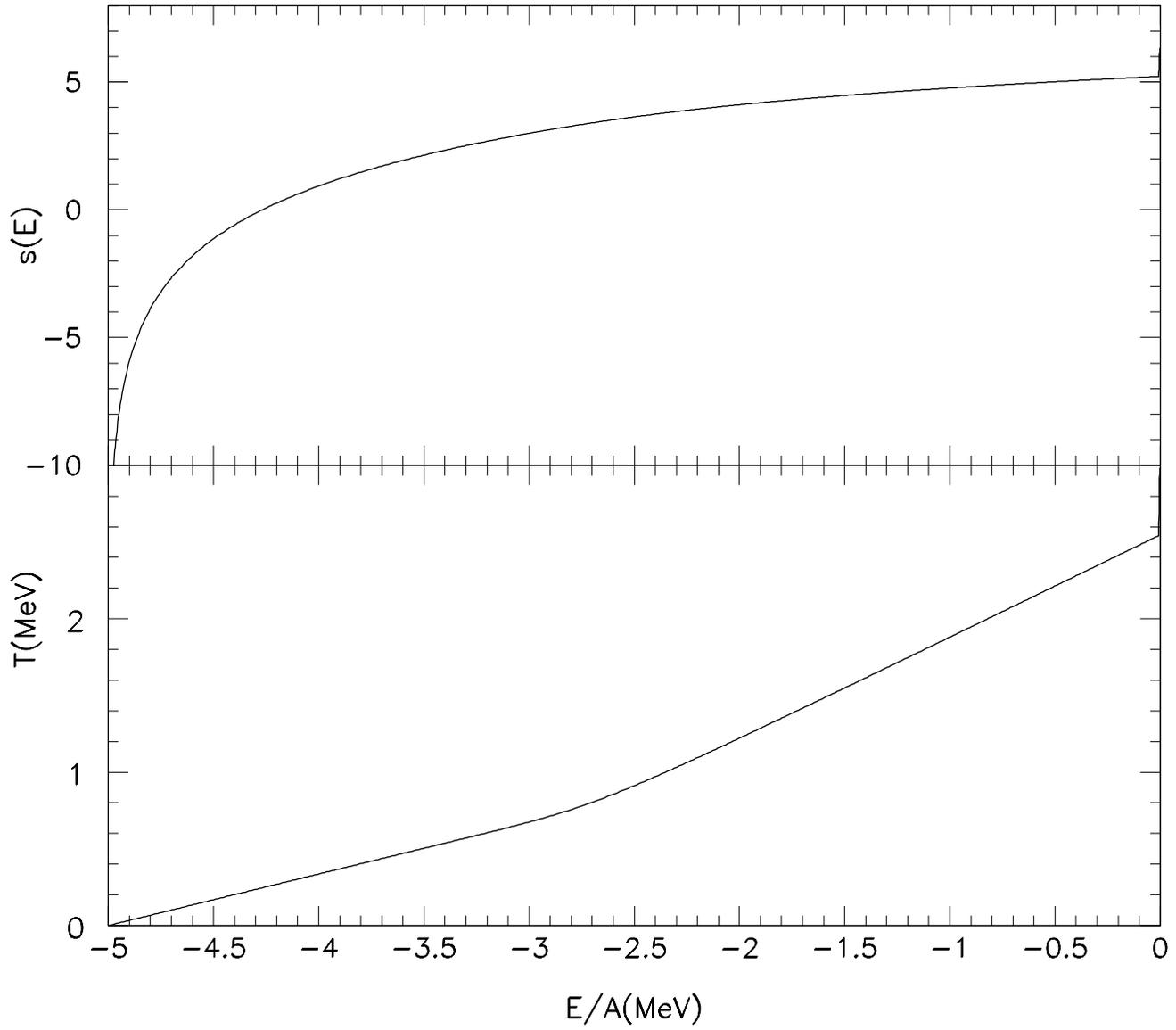
\bye